\documentstyle[11pt,epsfig]{article}
\newlength{\dinwidth}
\newlength{\dinmargin}
\begin{document}

\newcommand{\xgg}{x_{g}^{\gamma}}
\newcommand{\xgp}{x_{g}^{p}}
\newcommand{\xjg}{x_{j}^{\gamma}}
\newcommand{\xjp}{x_{j}^{p}}

\titlepage
\begin{flushright}
hep-ph/9903439 \\
DTP/99/28 \\
March 1999 \\
\end{flushright}

\begin{center}
\vspace*{2cm}
{\Large \bf Small $x$ QCD effects in DIS with a forward jet or a 
forward $\pi^{0}$.} \\

\vspace*{1cm}
J.\ Kwieci\'{n}ski$^{a,b}$, A.D.\ Martin$^{a}$ and J.J.\ Outhwaite$^{a}$.

\vspace*{0.5cm}

\begin{center}
{$^{a}$ Department of Physics, University of Durham, Durham, DH1 3LE, UK.}\\
{$^{b}$ H.\ Niewodniczanski Institute of Nuclear Physics, 
ul.\ Radzikowskiego 152, 31-342 Krakow, Poland.}
\end{center}
\end{center}
\vspace*{1.5cm}

\vspace*{2cm}

\begin{abstract}
We compare predictions based on small $x$ (QCD) dynamics with recent data
for deep inelastic events containing forward jets or forward $\pi^{0}$ 
mesons. We quantify the effect of imposing the (higher order) consistency 
condition on the BFKL equation and study uncertainties inherent in the QCD
predictions. We also estimate the cross-section for the forward production
of two jets.
\end{abstract}

\newpage
\section{Introduction}
The positron-proton collider at HERA in the DESY facility has opened a
window on a rich vein of fascinating physics. In recent years the range of
the available kinematic space for deep inelastic processes has
been extended ever further, and for a greater breadth of processes, allowing 
for increasingly stringent tests on 
the physics of Quantum Chromodynamics (QCD), the theory
of strong interactions, at perturbative energies.
Pertubative QCD allows us to predict the evolution in kinematic space 
(but not the initial form) of the parton distribution functions.
These in turn drive the equations that describe physical observables.
In distinct regions of $(x,Q^{2})$ space we have two different modes of 
parton evolution. At high $Q^{2}$ pQCD requires we resum the contributions of
$\alpha_{S} \log(Q^{2}/Q_{0}^{2})$ terms. This yields the well 
documented DGLAP equations. As centre of mass energy $\sqrt{s}$, increases
at moderate $Q^2$, we can technically
encounter a second large logarithm, this time in $1/x \sim s/Q^{2}$. 
Resummation of this type leads to the celebrated 
Balitskij-Fadin-Kuraev-Lipatov (BFKL) equation for the gluon.
The BFKL equation corresponds to ladder
diagrams with (reggeized) gluon exchange along the chain.
One of the main predictions of the BFKL formalism is singular power law
small $x$ behaviour, $x^{-\lambda}$, of deep inelastic scattering.
The exponent $\lambda$ has been calculated at leading 
order \cite{bfkl,lipatov}
and next-to-leading 
order \cite{nlo1,ross,salam} which contributes large negative 
corrections to the LO value.
The leading order approximation of the BFKL equation should therefore be
regarded as unreliable. 
It should be emphasised that the subleading $\log(1/x)$ effects are so
strong that their next-to-leading (NLO) approximation alone is entirely
unreliable already for reasonably small, yet relevant values, of the strong
coupling, that is for $\alpha_{S} > 0.15$. The exact form of the sub-leading
contribution, resummed to all orders, is unfortunately unknown. It may 
however be possible to pin down the dominant non-leading effects of 
well-defined physical origin, and to perform their exact resummation 
(that is going beyond the unreliable next-to-leading approximation)
\cite{salam,sutton1}.
In ref. \cite{sutton1} we have identified one class of such sub-leading corrections
which followed from imposing a certain kinematical consistency constraint
on the available phase space of the emitted gluons along the chain. At the 
next-to-leading level this constraint exhausts about 70\% of the corrections
to the BFKL exponent $\lambda$ and, more important, the effects of this 
constraint can be resummed to all orders. After including the constraint 
in the (LO) BFKL
equation, we obtain a formalism which makes it possible to implement
the dominant subleading effects and to resum them to all orders.
In ref. \cite{stasto} the BFKL equation supplemented by the consistency constraint
was used in a quantitative analysis of the structure function $F_{2}$ within
a unified BFKL and DGLAP scheme. In this paper we wish to study the effects
of these modifications on the properties of final state processes in deep 
inelastic scattering (DIS), which are particularly sensitive on BFKL
dynamics. These are DIS accompanied by a forward jet \cite{mueller} and 
the associated process of DIS with forward $\pi^{0}$'s \cite{sabine1}. 
The existing calculations 
of these processes have been performed within a LO BFKL framework, and it is
clearly essential to improve the analysis by incorporating the effects of the
above constraint.
The DIS + forward jet measurement first proposed by Mueller \cite{mueller}
can be a very 
useful tool for probing the BFKL dynamics where the diffusion of the 
transverse momentum along the gluon chain plays an important role. 
By choosing the configuration in which $k_{jT}^{2} \sim Q^{2}$, where
$k_{jT}^{2}$ denotes the jet transverse momentum, we eliminate the 
phase space region of strongly-ordered transverse momenta of the gluons 
along the chain, i.e. the region 
$k_{jT}^{2} \ll k_{1}^{2}... \ll k_{n}^{2} \ll Q^{2}$,
which is the dominant factor driving the increase of $F_{2}$ in the double leading
log approximation. Moreover in 
the forward configuration $\xjp \gg x$, where $x$ is the Bjorken parameter,
we have large subenergy available for jet production that justifies the use
of BFKL dynamics.
We can also study DIS + forward $\pi^{0}$'s \cite{sabine1}. This is a more
refined version of the DIS + forward jet case, 
in the sense that we no longer have
potential ambiguities derived from hadronisation effects or jet finding 
algorithms. The distinguishing features are discussed more fully in 
section 5.
We first motivate higher order constraints which should naturally be
implemented in a BFKL formalism.
We then show how observables for DIS + forward jet/$\pi^{0}$  can be
calculated from the unintegrated gluon distribution, evolved from the 
virtual photon end of the BFKL ladder
\footnote{The present form of analysis in which we 
iterate from the virtual photon will 
facilitate comparison with the analysis of forward jet production in terms of the
partonic structure of the virtual photon.}
, unlike the hybrid
evolution used in ref. \cite{sabine1}. It is more convienient to impose the
higher order constraints in the present form of the evolution.
We then compare our calculations
with experimental results for DIS + forward jet production
from H1 \cite{h1} and ZEUS \cite{zeus}, and show a 
comparison of our analysis with very recent results for $\pi^{0}$
cross-sections obtained by the H1 collaboration at HERA \cite{new_h1}. 
Finally we estimate the cross-section
for production of two jets satisfying the forward criteria.
\section{The unintegrated BFKL gluon from $\gamma^{*} g$ fusion}
In this section we shall describe the formalism needed for the theoretical 
description of the DIS + forward jet measurement based on the BFKL equation
\cite{sutton2}.
\begin{figure}[h]
\label{fig:glu}
\begin{center}
\mbox{\epsfig{figure=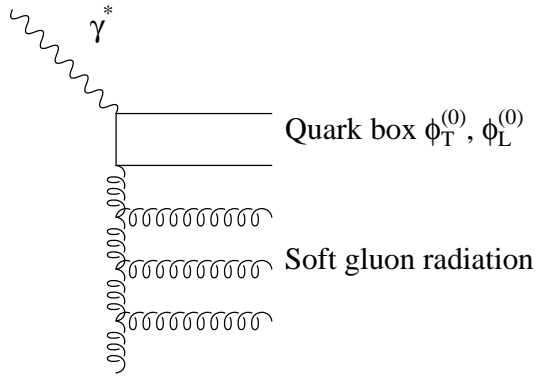,width=10cm}}
\caption{Diagram showing the $\gamma^{*} g$ fusion, including 
the soft gluon radiation ladder.}
\end{center}
\end{figure}
The natural quantity driving physical processes in DIS in the small Bjorken-$x$ 
region of $(x,Q^{2})$ phase-space is the {\it unintegrated gluon distribution},
$\phi_{i}$
\begin{equation}
\xgg g_{i}(\xgg, Q^{2},\bar{Q^{2}}) = \int^{\bar{Q^{2}}}
{dk_{T}^{2} \over k_{T}^{2}} \phi_{i}(\xgg,k_{T}^{2},Q^{2}),
\end{equation}
where $g_{i}(\xgg,Q^{2},\bar{Q^{2}})$ are the conventional gluon 
distributions of a photon of virtuality $Q^{2}$ and polarisation $i = T, L$,
probed at a scale $\bar{Q^{2}}$.
The parameter $\xgg$ denotes
the longitudinal momentum fraction of the parent virtual photon carried by
the gluon and $k_{T}$ denotes its transverse momentum. The gluon four momentum
$k$ has the following Sudakov decomposition
\begin{equation}
k = -\xgp p + \xgg q^{\prime} + {\mbox{\boldmath${k_{T}}$}}, 
\label{eqn:sudakov}
\end{equation}
where $p$ is the four momentum of the nucleon, and the light-like vector
$q^{\prime}$ is defined by 
\begin{equation}
q^{\prime} = q + x p.
\end{equation}
The variable $x$ is the Bjorken parameter conventionally defined as
$x = Q^{2}/(2p.q)$ and, as usual, $Q^{2} = -q^{2}$ with $q$ being the
four momentum carried by the virtual photon.
The contribution to $\phi$ arising from the virtual photon can be thought of
in terms of virtual photon-gluon fusion through a {\it quark-box} coupled
to an equation describing the subsequent evolution of the gluon distribution.
The quark box driving terms, $\phi_{T}^{(0)}$ and $\phi_{L}^{(0)}$,
can be evaluated perturbatively
\begin{displaymath}
\phi_{T}^{(0)}(\xgg,k_{T}^{2},Q^{2}) = 
\end{displaymath}
\begin{equation}
\sum_{q} e_{q}^{2}
{ \alpha_{S} \over 4 \pi^{2} }
{ Q^{2} } \int_{\xgg}^{1} d\beta \int d^{2}
{\mbox{\boldmath${\kappa_{T}}$}}
\left\{ \left[ \beta^{2} + (1-\beta)^{2} \right] \left( 
{ {\mbox{\boldmath${\kappa_{T}}$} } \over D_{1} } -
{ { ( {\mbox{\boldmath${\kappa_{T}}$}} - {\mbox{\boldmath${k_{T}}$}} ) } 
\over D_{2} } \right)^{2} + m_{q}^{2}
\left( {1 \over D_{1} } - { 1 \over D_{2} } \right)^{2} \right\},
\label{eqn:qbox}
\end{equation}
\begin{displaymath}
\phi_{L}^{(0)}(\xgg,k_{T}^{2},Q^{2}) = \sum_{q} e_{q}^{2} { \alpha_{S} \over \pi^{2} }
{ Q^{2} } \int_{\xgg}^{1} d\beta \int d^{2} {\mbox{\boldmath${\kappa_{T}}$}}
\beta^{2}(1-\beta)^{2} \left( {1 \over D_{1}} - {1 \over D_{2}} \right)^{2}
\end{displaymath}
where the denominator functions are
\begin{displaymath}
D_{1} = \kappa_{T}^{2} + \beta(1-\beta)Q^{2} + m_{q}^{2},
\end{displaymath}
\begin{displaymath}
D_{2} = (
{\mbox{\boldmath${\kappa_{T}}$}}
- {\mbox{\boldmath${k_{T}}$} })^{2} 
+ \beta(1-\beta)Q^{2} + m_{q}^{2}.
\end{displaymath}
The quark mass $m_{q}$ is set to zero for the light $u,d,$ and $s$ quarks
and taken to be $1.4$GeV for the charm quark.
Taking these driving terms we then evolve the unintegrated gluon 
distributions, $\phi_{T}$ and $\phi_{L}$,
using the BFKL equation from the virtual 
photon end of the BFKL ladder
\begin{displaymath}
\phi_{i}(\xgg, k_{T}^{2}, Q^{2}) = \phi_{i}^{(0)}(\xgg, k_{T}^{2}, Q^{2}) +
\end{displaymath}
\begin{displaymath}
\bar{\alpha_{S}}(k_{T}^{2}) k_{T}^{2} 
\int_{\xgg}^{1} {dz \over z} \int_{k_{0}^{2}}^{\infty}
{d k_{T}^{\prime 2} \over k_{T}^{\prime 2} }
\left\{ { 
\phi_{i}( {\xgg/z}, k_{T}^{\prime 2}, Q^{2} ) - \phi_{i}
( {\xgg/z}, k_{T}^{2}, Q^{2} )
\over \mid k_{T}^{2} - k_{T}^{\prime 2} \mid }
+ { \phi_{i}( {\xgg/z}, k_{T}^{2}, Q^{2} ) \over
\left( 4 k_{T}^{\prime 4} + k_{T}^{4} \right)^{ 1 \over 2 } } \right\},
\end{displaymath}
\begin{equation}
\label{eqn:bfkl}
\end{equation}
where $i = T$ or $L$, $\bar{\alpha_{S}} = 3{\alpha_{S}}/\pi$, 
and $\xgg$ is the fraction of the virtual photon's longitudinal
momentum carried by the gluon. We will show results for 
infrared cut-off $k_{0}^{2}$ varied 
within the range $0.5 - 1$GeV$^{2}$.
We require strong ordering in  $(\xgg)_{i}$ along the gluon chain
\begin{displaymath}
(\xgg) \ll (\xgg)_{1} \ll ... \ll (\xgg)_{n}.
\end{displaymath}
The present approach differs from our earlier work 
\cite{sabine1} in that we no longer 
have a free parameter $z_{0}$ which 
specifies the end of the BFKL evolution. 
In our previous treatment \cite{sabine1} we performed a hybrid 
evolution of the gluon ladder
in the sense that, while we allowed the $k_{T}^{2}$ diffusion from the quark box,
the longitudinal momentum fractions were defined with respect to the proton end
of the ladder and strongly ordered along the gluon chain.
In that case  $z_{0}$ was 
adjusted to give the correct normalisation for the DIS + jet data. 
In the present calculation the 
normalisation is, in principle, fixed by the theory. However,
in practice, there is freedom due to the choice of the value of 
QCD scale, which we will quantify later. 
In summary, the present description of the behaviour of 
the quark-box gluon chain system
can be considered as a calculation of the unintegrated gluon content 
of the virtual photon.
\section{Higher order corrections}
Calculations \cite{nlo1,ross,salam} have shown that the NLO corrections 
to the BFKL equation
are large. This casts doubt upon the quantitative predictive power of any
LO BFKL approach to small $x$ DIS dynamics. Clearly, for phenomenological
purposes, a method of including the higher order effects should be considered.
One origin of subleading effects, which contributes to all orders, is the 
so called {\it consistency constraint} (CC) \cite{sutton1}.
This requires that the virtuality of the emitted gluons 
along the chain should arise 
predominantly from the transverse components of momentum, in order for the
small $x$ (small $\xgg$) approximation to be valid, that is
\begin{equation}
q_{T}^{2} < {k_{T}^{2} \over z},
\end{equation}
where we have omitted a factor of $(1-z)$.
We can motivate the inclusion of the consistency constraint in the 
analysis by reference to the exponent $\lambda$ governing small $x$ behaviour, 
$x^{-\lambda}$. It has been shown \cite{sutton1} that by 
truncating the all order BFKL + CC
solution at NLO, one recovers some $70\%$ of the full, explicit NLO calculation
in the effective exponent.
It looks reasonable to suppose that this comprises the 
dominant part of all higher
order effects. 
Another physical source of sub-leading contributions comes from the imposition
of the conservation of energy-momentum in multigluon emission. Such an 
effect has been investigated in ref. \cite{wjs}. It turns out that the 
consistency constraint subsumes energy-momentum conservation over a wide
region of the allowed phase space \cite{sutton1}.
\begin{figure}[h]
\label{fig:cc}
\begin{center}
\mbox{\epsfig{figure=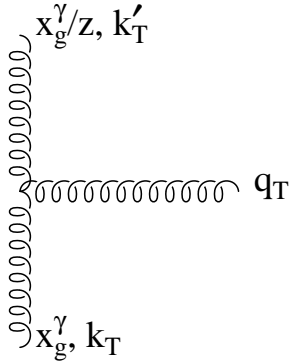,width=4cm}}
\end{center}
\caption{Diagram to illustrate the radiation of soft gluons from the 
reggeized gluonic propagator in the {\it t}-channel.}
\end{figure}
The CC places restrictions on the available phase space under the integration
in the BFKL equation (\ref{eqn:bfkl}).
This manifests itself as a $\Theta$-function multiplying the kernel component
governing real gluon emissions. Upon implementing this, we obtain a modified BFKL
equation
\begin{displaymath}
\phi_{i}(\xgg, k_{T}^{2}, Q^{2}) = \phi_{i}^{(0)}(\xgg, k_{T}^{2}, Q^{2}) +
\end{displaymath}
\begin{displaymath}
\bar{\alpha_{S}} k_{T}^{2} \int_{\xgg}^{1} {dz \over z} \int_{k_{0}^{2}}^{\infty}
{d k_{T}^{\prime 2} \over k_{T}^{\prime 2} }
\left\{ { \Theta \left( {k_{T}^{2}/k_{T}^{\prime 2}} - z \right)
\phi_{i}( {\xgg/z}, k_{T}^{\prime 2}, Q^{2} ) - \phi_{i}
( {\xgg/z}, k_{T}^{2}, Q^{2} )
\over \mid k_{T}^{2} - k_{T}^{\prime 2} \mid }
+ { \phi_{i}( {\xgg/z}, k_{T}^{2}, Q^{2} ) \over
\left( 4 k_{T}^{\prime 4} + k_{T}^{4} \right)^{ 1 \over 2 } } \right\}.
\end{displaymath}
\begin{equation}
\label{eqn:mbfkl}
\end{equation}
\section{Forward jet production in DIS.}
\begin{figure}[h]
\label{fig:bfkl}
\begin{center}
\mbox{\epsfig{figure=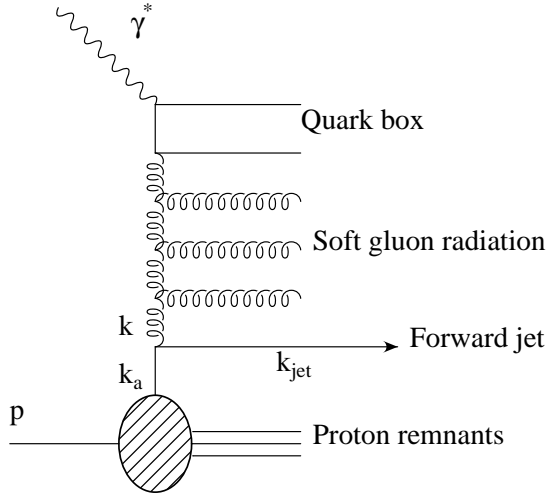,width=8cm}}
\caption{Diagram showing forward jet production driven by $\gamma^{*} g$
fusion coupled to evolution through a BFKL-type ladder. The struck parton
from the proton could in principle be either a gluon or a quark.}
\end{center}
\end{figure}
Given that measurements of $F_{2}$ proved unable to unambiguously
determine small $x$ dynamics, less inclusive alternative approaches were needed.
One such process is Mueller's proposal \cite{mueller}: 
\begin{displaymath}
\gamma^{*} + p \longrightarrow \mathrm{jet} + {\it X}
\end{displaymath}
According to Mueller, DIS events containing an identified forward jet provide a
particularly clean window on small $x$ dynamics. By requiring that 
$Q^{2} \simeq k_{jT}^{2}$ we ensure minimal DGLAP
type evolution, and any remaining features must be the result of
dynamics from the small $x$ limiting region of the $(x,Q^{2})$ phase-space.
Moreover by considering jets with large fractions of the proton's
longitudinal momentum, $\xjp$ allowing $x/\xjp$ to be as small as possible,
on the one hand, will allow the use of BFKL dynamics, while on the other hand, 
involve parton
distributions at values of $\xjp$ where they are well known.
\subsection{QCD prediction for DIS + forward jet}
The kinematics are such that we may use
{\it strong ordering} at the parton-gluon-jet vertex 
\begin{displaymath}
k_{aT}^{2} \ll k_{jT}^{2}, \; \; 
\xjp \gg x,
\end{displaymath}
where the Sudakov decomposition of the jet four momentum in terms of light-like momenta
$p$ and $q^{\prime}$ in analogy to (\ref{eqn:sudakov}) is 
\begin{equation}
k_{j} = \xjp p + \xjg q^{\prime} + {\mbox{\boldmath${k_{jT}}$}}. 
\end{equation}
We can relate the longitudinal momentum fractions at the jet vertex through 
strong-ordering and the jet on-shell condition, $k_{j}^{2} = 0$. 
This allows us to deduce that
\begin{displaymath}
2\xjp \xjg (p.q^{\prime}) = k_{jT}^{2},
\end{displaymath}
\begin{displaymath}
\xgg \simeq \xjg = {k_{jT}^{2} \over 2\xjp (p.q)} = 
{k_{jT}^{2} x \over Q^{2} \xjp}.
\end{displaymath}
Using the prescription described in the previous section,
we solve numerically for $\phi$ in the modified BFKL equation, via an 
expansion in Chebyshev polynomials, and finally determine the unintegrated 
gluon distribution
\begin{equation}
\Phi_{i}( { x \over \xjp }, k_{jT}^{2}, Q^{2} ) \; = \;
\phi_{i}( \xgg = { k_{jT}^{2} x \over Q^{2} \xjp }, k_{jT}^{2}, Q^{2} ).
\end{equation}
$\Phi_{i}( { x/\xjp }, k_{jT}^{2}, Q^{2} )$ can then be used to calculate
the differential structure functions that drive the forward jet process through
\begin{displaymath}
{ \partial^{2} F_{i} \over \partial \xjp \partial k_{jT}^{2} } =
{3 \alpha_{S} \over \pi k_{jT}^{4} }
\left( \sum_{a} f_{a}(\xjp,k_{jT}^{2}) \right) \Phi_{i}
\left({x \over \xjp}, k_{jT}^{2},Q^{2} \right),
\end{displaymath}
\begin{equation}
\label{eqn:partons}
\end{equation}
where $\sum_{a} f_{a}$ is the sum over parton distributions 
assuming {\it t}-channel pole dominance.
\begin{equation}
\sum_{a} f_{a} = g + {4 \over 9}\sum_{q}(q + \bar q).
\label{eqn:tpole}
\end{equation}
These quantities can be substituted into an expression for the differential 
cross-section for forward jets
\begin{equation}
{ \partial \sigma_{j} \over
\partial x \partial Q^{2} \partial \xjp \partial k_{jT}^{2} } =
{ 4 \pi \alpha^{2} \over x Q^{4} } \left[ (1-y)
{\partial^{2} F_{2} \over \partial{\xjp} \partial k_{jT}^{2} } + {1 \over 2}y^{2}
{\partial^{2} F_{T} \over \partial{\xjp} \partial k_{jT}^{2} } \right],
\label{eqn:xs}
\end{equation}
where 
$y = { (p.q) / (p_{e}.q) }$
 and
$F_{2} = F_{T} + F_{L}$.\\
\subsection{Comparison with HERA data}
Finally we are in a position to be able to confront our calculation
with the HERA data. We numerically integrate the differential 
cross-section over the kinematic regions used by the H1 \cite{h1} 
and ZEUS \cite{zeus} collaborations, see Table~\ref{tab:cuts1}.
The results are shown in Fig.4. 
Since the parton distributions (\ref{eqn:tpole}) are required in a kinematic domain 
where they are well known the results are not sensitive to the particular
set that is used. However there is a dependence of the results on the
QCD scale. To be consistent we use a recent {\it leading order} set of 
partons. We take the LO set from \cite{mrst}, for which $\Lambda({\mathrm{QCD}})
= 0.174$GeV for four flavours. 
The three curves in  the figure correspond to the arguments of $\alpha_{S}$ 
in (\ref{eqn:qbox}) and (\ref{eqn:partons}) and the infrared cut-off 
in (\ref{eqn:mbfkl}) being respectively taken to be 
\footnote{When the scale of $\alpha_{S}$ is less than $k_{0}^{2}$, in choices
(i) and (ii), we freeze the coupling at $\alpha_{S}(k_{0}^{2})$.}
\begin{equation}
\begin{array}{ccccc}
\mathrm{(i)} & (k_{T}^{2} + m_{q}^{2})/4, & k_{T}^{2}/4, 
& k_{0}^{2} = 0.5\mathrm{GeV}^{2}
& \mathrm{(upper \; dashed)} \\
\mathrm{(ii)} & (k_{T}^{2} + m_{q}^{2})/4, & k_{T}^{2}/4, 
& k_{0}^{2} = 1\mathrm{GeV}^{2}
& \mathrm{(continuous)} \\
\mathrm{(iii)} & (k_{T}^{2} + m_{q}^{2}), & k_{T}^{2}, 
& k_{0}^{2} = 0.5\mathrm{GeV}^{2}
& \mathrm{(lower \; dashed)}.
\end{array}
\label{eqn:scales}
\end{equation}
The sensitivity to the choice of scales for $\alpha_{S}$ is therefore
seen by comparing the predictions for (i) and (iii), and the sensitivity
to the value chosen for the infrared cut-off $k_{0}^{2}$ by comparing (i) 
and (ii). 
We see that the uncertainty due to $k_{0}^{2}$ is much less than the 
uncertainty due to the choice of scales.
From Fig. 4 we see that the shape of the $x$ distributions of
the DIS + forward jet data are modelled well. Moreover the predicted 
normalisation is satisfactory in that agreement exists for a physically 
reasonable choice of scales and of the infrared cut-off $k_{0}^{2}$.
The experimental data correspond to the measurements at the hadron level. 
The hadronisation effects can lower the cross-section by about 15-20\%.
We comment on the comparison with data 
further in section 7.
\begin{table}[h]
\begin{center}
\begin{tabular}{|c|c|} \hline
H1 cuts & ZEUS cuts\\ \hline \hline
$E_{e}^{\prime} > 11$GeV & $E_{e}^{\prime} > 10$GeV \\
$y_{e} > 0.1$ & $y_{e} > 0.1$ \\
$\xjp > 0.035$ & $\xjp > 0.036$ \\
$k_{jT} > 3.5$GeV & $E_{jT} > 5$GeV \\
$0.5 < k_{jT}^{2} / Q^{2} < 2$ & $0.5 < E_{jT}^{2} / Q^{2} < 2$ \\
$7^{o} < \theta_{jet} < 20^{o}$ & $\eta_{jet} < 2.6$ \\
$160^{o} < \theta_{e} < 173^{o}$ & \\ \hline
\end{tabular}
\end{center}
\caption{Table showing the kinematic restrictions imposed on DIS events at HERA
for forward jet production.}
\label{tab:cuts1}
\end{table}
\begin{figure}
\begin{center}
\mbox{\epsfig{figure=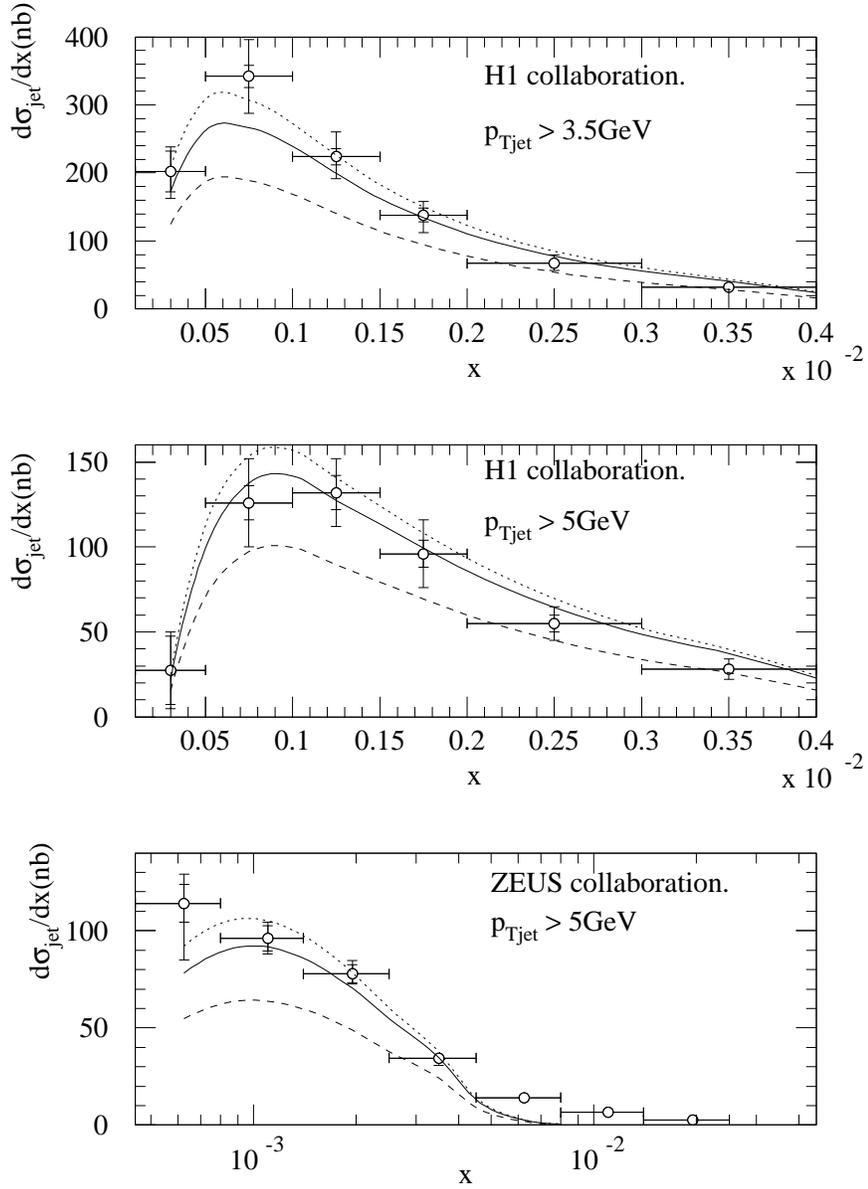,width=12cm}}
\caption{The DIS + forward jet differential cross-section 
versus Bjorken-$x$ as measured at the hadron level by the H1 \cite{h1}
and ZEUS \cite{zeus} 
collaborations. 
The kinematic cuts are 
given in Table~\ref{tab:cuts1}.
The curves are predictions at the parton level, 
based on the BFKL formalism including sub-leading
corrections, corresponding to the three choices of scales and infrared
cut-off given in (\ref{eqn:scales})
.}
\label{fig:fj94}
\end{center}
\end{figure}
\section{DIS events containing a forward $\pi^{0}$}
\begin{figure}[h]
\label{fig:eps}
\begin{center}
\mbox{\epsfig{figure=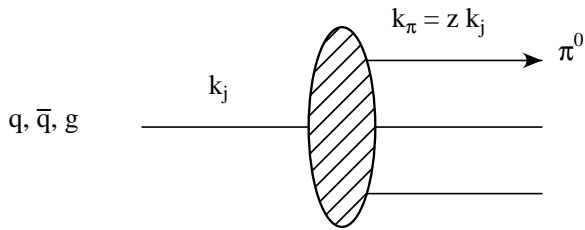,width=8cm}}
\end{center}
\caption{Schematic showing collinear fragmentation of a forward parton jet 
of momentum $k_{j}$ into a forward pion of momentum $z k_{\pi}$.
The process is described by fragmentation functions, $D_{i}(z,\mu^{2})$}
\end{figure}
A complementary reaction to Mueller's forward jet process is provided
by the production of forward $\pi^{0}$ mesons in deep inelastic events
\cite{sabine1}. The driving process here is the same 
as before, with
$\gamma^{*}g$ fusion coupled to BFKL-type gluon evolution, ejecting a parton
within the proton as a forward jet. However, we now allow the parton to 
evolve into a shower of particles containing a pion collinear to the
initial parton jet
\begin{displaymath}
\gamma^{*} + p \longrightarrow \pi^{0} + X.
\end{displaymath}
There are advantages in attempting to measure the 
forward $\pi^{0}$ cross-section as compared to the parent forward-jet
cross-section.
\begin{itemize}
	\item{Experimentally one can more unambiguously identify a forward
	$\pi^{0}$ than a jet.}
	\item{By measuring $\pi^{0}$'s at relatively low $x_{\pi}, p_{T\pi}$
	we effectively collect data for energetic forward jets with
	$x_{j} > x_{\pi}$ and $k_{jT} > p_{T\pi}$, which might otherwise
	go undetected.}
	\item{Another consequence of looking for a specific final state
	is that we eliminate our dependence on jet-finding algorithms.
	We can unambiguously determine the signal due to the measurement
	of a real $\pi^{0}$ instead of worrying about jet definitions.}
	\item{Finally, we eliminate hadronisation uncertainties.
	All non-perturbative hadronisation effects are swept into the
	fitted parametrisations of the fragmentation functions.}
\end{itemize}
On the other hand, by requiring a single energetic fragment of the jet
we reduce the cross-section.
\subsection{Including forward jet $\rightarrow \pi^{0}$ fragmentation}
We use the LO $\pi^{\pm}$ fragmentation functions 
of Binnewies et al. \cite{binnewies}. They present results of the form
\begin{equation}
D_{i}(z,\mu^{2}) = N_{i}(\mu^{2}) z^{\alpha_{i}(\mu^{2})}
(1-z)^{\beta_{i}(\mu^{2})}
\label{eqn:frag}
\end{equation}
where $i = g, q, \bar{q}$ and $\mu^{2}$ is the fragmentation scale. Here 
we take the fragmentation scale simply as $k_{jT}^{2}$.
The functions for $\pi^{0}$ are taken to be ${ 1 \over 2}(\pi^{+} + \pi^{-})$ 
distributions
The form of the fragmentation requires the $\pi^{0}$ to carry 
$ z = x_{\pi}/\xjp $ of the jet's longitudinal momentum, in a direction
collinear with the initial parton.
With this assumption we can relate the transverse momentum of the $\pi^{0}$
and the forward jet
\begin{displaymath}
\mbox{\boldmath${k_{\pi T}}$} = z \mbox{\boldmath${k_{j T}}$}.
\end{displaymath}
Next we obtain the cross-section for $\pi^{0}$ production by 
convoluting the DIS + forward jet
cross-section (\ref{eqn:xs})
with the $\pi^{0}$ fragmentation parametrisations 
(\ref{eqn:frag}) \cite{sabine1}
\begin{displaymath}
{ \partial\sigma_{\pi^{0}} \over \partial x_{\pi} \partial p_{T\pi}^{2}
\partial x \partial Q^{2} } =
\int_{x_{\pi}}^{1} dz \int d\xjp \int dk_{jT}^{2} \delta( x_{\pi} - z\xjp ) 
\delta( p_{T\pi} - zk_{jT})
\end{displaymath}
\begin{displaymath}
\left\{
{ \partial \sigma_{g} \over \partial \xjp \partial k_{jT}^{2} \partial x
\partial Q^{2} } D_{g}^{\pi^{0}}(z,k_{jT}^{2}) + {4 \over 9} \sum_{q} \left[
{ \partial \sigma_{q} \over \partial \xjp \partial k_{jT}^{2} \partial x
\partial Q^{2} } D_{q}^{\pi^{0}}(z,k_{jT}^{2}) +
{ \partial \sigma_{\bar q} \over \partial \xjp \partial k_{jT}^{2} \partial x
\partial Q^{2} } D_{\bar q}^{\pi^{0}}(z,k_{jT}^{2}) \right] 
\right\}.
\end{displaymath}
\begin{equation}
\end{equation}
The differential cross-sections on the right-hand-side, 
$\partial \sigma_{j} /\partial \xjp \partial k_{jT}^{2} \partial x \partial Q^{2}$
with $j= q,{\bar{q}}, g,$ are given by (\ref{eqn:partons}) and (\ref{eqn:xs}) 
with the replacement of
$\sum f_{a}$ by $f_{q}, f_{\bar{q}}$ and $f_{g}$ respectively.
\subsection{Comparison with forward $\pi^{0}$ data}
\begin{figure}
\begin{center}
\mbox{\epsfig{figure=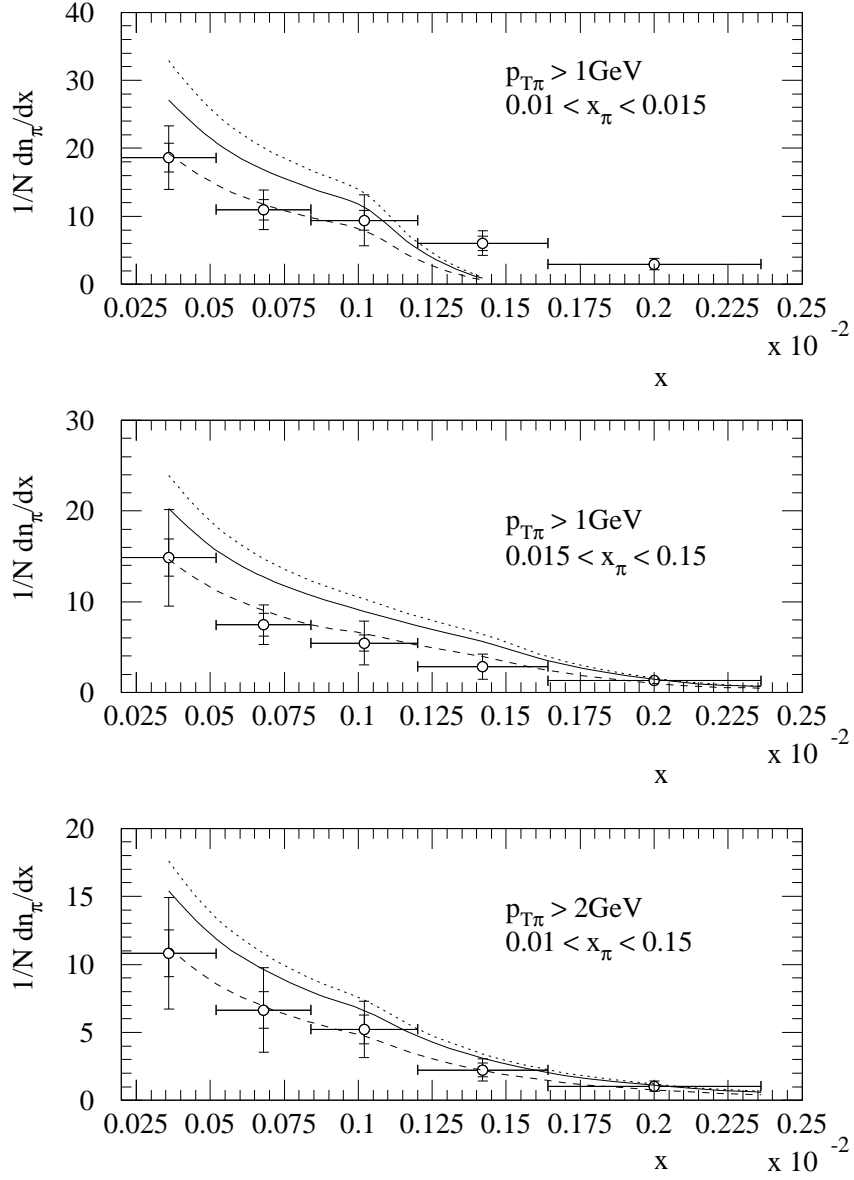,width=12cm}}
\caption{The $\pi^{0}$ spectra (\ref{eqn:spectra}) 
versus Bjorken-$x$ obtained from 1994 H1 data \cite{h1}.
The curves are predictions, based on the BFKL formalism including sub-leading
corrections, corresponding to the three choices of scales and infrared
cut-off given in (\ref{eqn:scales}).
[The restriction $x/x_{\pi} < 0.1$ limits the comparison to the domain
$x \lower .7ex\hbox{$\;\stackrel{\textstyle<}{\sim}\;$} 10^{-3}$ 
in the upper plot.]
}
\label{fig:h1}
\end{center}
\end{figure}

There exist two sets of measurement of deep inelastic events containing
forward $\pi^{0}$ mesons. First there are the 1994 H1 data \cite{h1}, and now the 
recent analysis of the 1996 H1 data \cite{new_h1}. The kinematic cuts used to obtain 
these data samples are summarized in Table~\ref{tab:cuts2}.
The $\pi^{0}$ spectra are defined through
\begin{equation}
{1 \over N}{ dn_{\pi} \over dx} = 
{1 \over \sigma_{tot} }
{ \partial \sigma_{\pi} \over \partial x} 
\label{eqn:spectra}
\end{equation}
where $n_{\pi}$ is the number of $\pi^{0}$'s, and $N$ is the total number of DIS
events.
The recent H1 results \cite{new_h1} are fully comprehensive, 
with histograms showing the 
differential cross-section for $\pi^{0}$ production as functions of 
$x$, $Q^{2}$, $p_{T\pi}$ and $\eta_{\pi}$.
We compute the $\pi^{0}$ cross-section using exactly the same three choices
of the scales of $\alpha_{S}$ and infrared cut-off $k_{0}^{2}$ as we used for 
the DIS + forward jet predictions, see Section 4.2.
\begin{table}[h]
\begin{center}
\begin{tabular}{|c|c|} \hline
1994 H1 data & New H1 data\\ \hline \hline
$E_{e}^{\prime} > 12$GeV & \\
$y_{e} > 0.1$ & $ 0.6 > y_{e} > 0.1$ \\
$x_{\pi} > 0.01$ & $x_{\pi} > 0.01$ \\
$p_{T\pi} > 1$GeV & $p_{T\pi} > 2.5$GeV\\
$5^{o} < \theta_{\pi} < 25^{o}$  & $5^{o} < \theta_{\pi} < 25^{o}$ \\
$156^{o} < \theta_{e} < 173^{o}$ & \\ 
& $2 < Q^{2} < 70$GeV$^{2}$\\ \hline 
\end{tabular}
\end{center}
\caption{Table showing the kinematic restrictions imposed on DIS events
by the H1 collaboration
for forward $\pi^{0}$ production. For the 1996 data, implicit bounds are imposed on
$\theta_{e}$ and $E_{e}^{\prime}$ by the $y_{e}$ restriction.}
\label{tab:cuts2}
\end{table}
In Fig.~\ref{fig:h1} we compare our analysis with the 1994 H1 $\pi^{0}$
spectra in three $x_{\pi}$ regimes. We find a satisfactory description of the 
shape, and that the normalisation is best described by choice (iii) of
the scales of (\ref{eqn:scales}).
The factor of two overshoot of the data
by the previous predictions \cite{sabine1} is gone. The reason is the 
suppresion due to the inclusion of the sub-leading corrections.
In Figures 7 through 10 we compare our calculations
of the $\pi^{0}$ differential cross-sections as a function of 
$x$, $Q^{2}$, $p_{T\pi}$ and $\eta_{\pi}$, with the new H1 data \cite{new_h1}.
As in the DIS + forward jet process, we see that there is good overall 
agreement between the predictions and the DIS + forward $\pi^{0}$ data.
The continuous curves, which are the set which best describe the
forward jet data, are on average some 20\% above the forward $\pi^{0}$ data.
However the overall agreement is well within the theoretical and
experimental uncertainties. (Note that no allowance has been made for the
uncertainty in the fragmentation functions for the pion).
Moreover note that the jet data are shown at the hadron level. If we were to correct
for hadronisation effects, then the parton level jet data are expected to be 15-20\%
lower \cite{wengler_pc}, and the overall consistency of the QCD description is
even better than shown.
If we look at the description of the forward $\pi^{0}$ data in more detail
then we see, from Figs. 7 and 9,
that the $Q^{2}$ behaviour of the DIS + 
forward $\pi^{0}$ data is not reproduced in detail by any single one of
the three sets of predictions. For instance for $Q^{2} < 4.5$GeV$^{2}$
the continuous curve is below the data, although in agreement within the 
errors, whereas for $Q^{2} > 4.5$GeV$^{2}$ the curve is above the data
by about two standard deviations on average. Note that the relative $Q^{2}$
behaviour evident between Figs. 7(b,c,d) simply reflects the comparison of 
the curves with the data in Fig. 9(a).
However, it is worth repeating that the overall agreement between the 
predictions of small $x$ dynamics and the DIS + $\pi^{0}$ data is much 
improved since the previous comparison \cite{sabine1}.
We comment further on the predictions
in Section 7.
\begin{figure}
\label{fig:d_bjorken}
\begin{center}
\mbox{\epsfig{figure=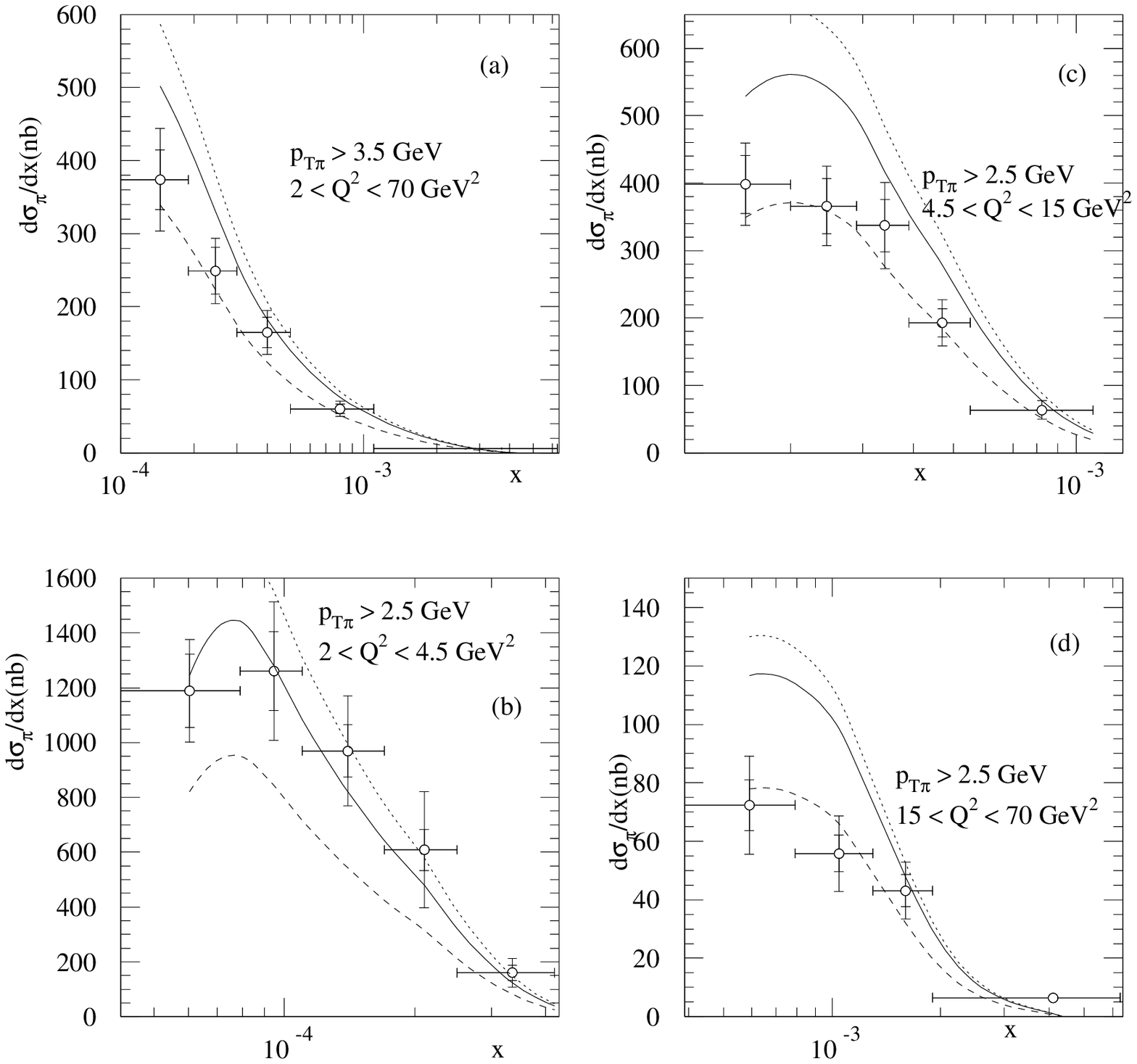,width=12cm}}
\caption{The $\pi^{0}$ differential cross-section versus
Bjorken-$x$ obtained from 1996 H1 
(preliminary) data \cite{new_h1}.
The curves are predictions, based on the BFKL formalism including sub-leading
corrections, corresponding to the three choices of scales and infrared
cut-off given in (\ref{eqn:scales})
.}
\end{center}
\end{figure}
\begin{figure}
\label{fig:d_ptpi0}
\begin{center}
\mbox{\epsfig{figure=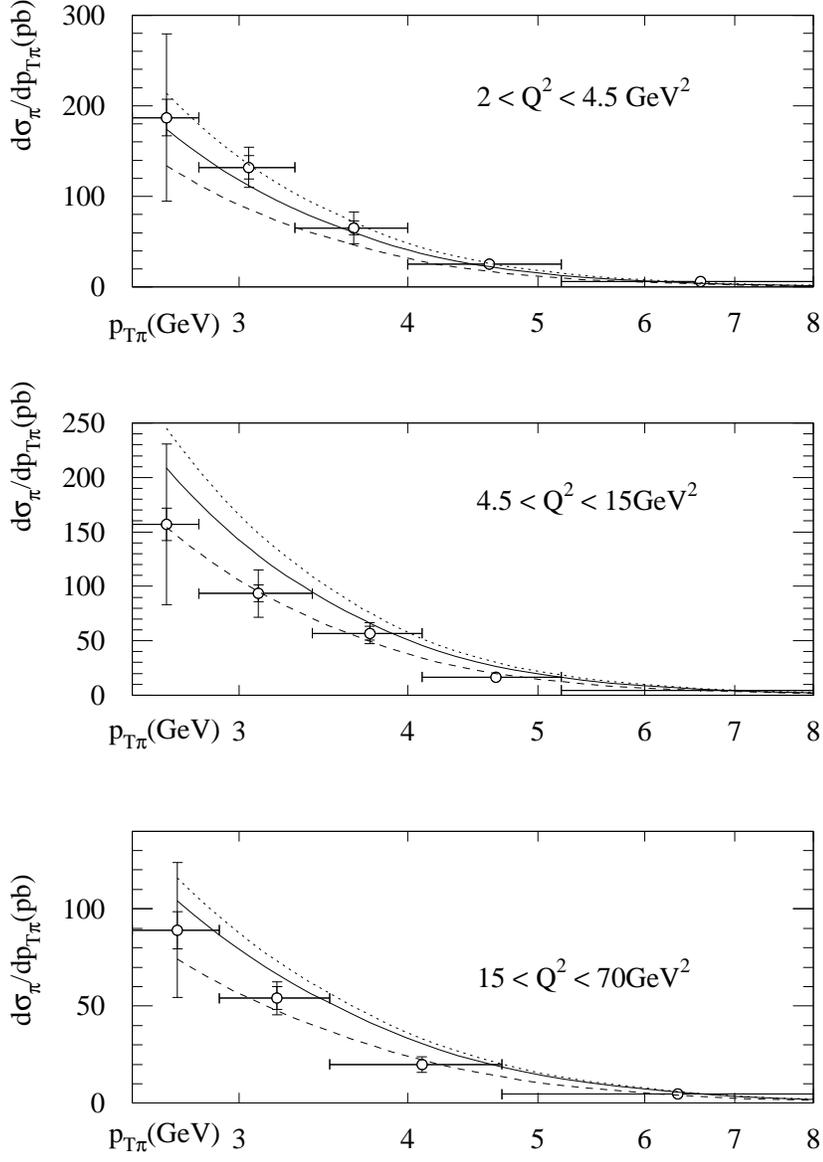,width=12cm}}
\caption{$\pi^{0}$ differential cross-section in transverse momentum,
obtained from 1996 H1 (preliminary) data \cite{new_h1}.
The curves are predictions, based on the BFKL formalism including sub-leading
corrections, corresponding to the three choices of scales and infrared
cut-off given in (\ref{eqn:scales})
.}
\end{center}
\end{figure}
\begin{figure}
\label{fig:d_q2}
\begin{center}
\mbox{\epsfig{figure=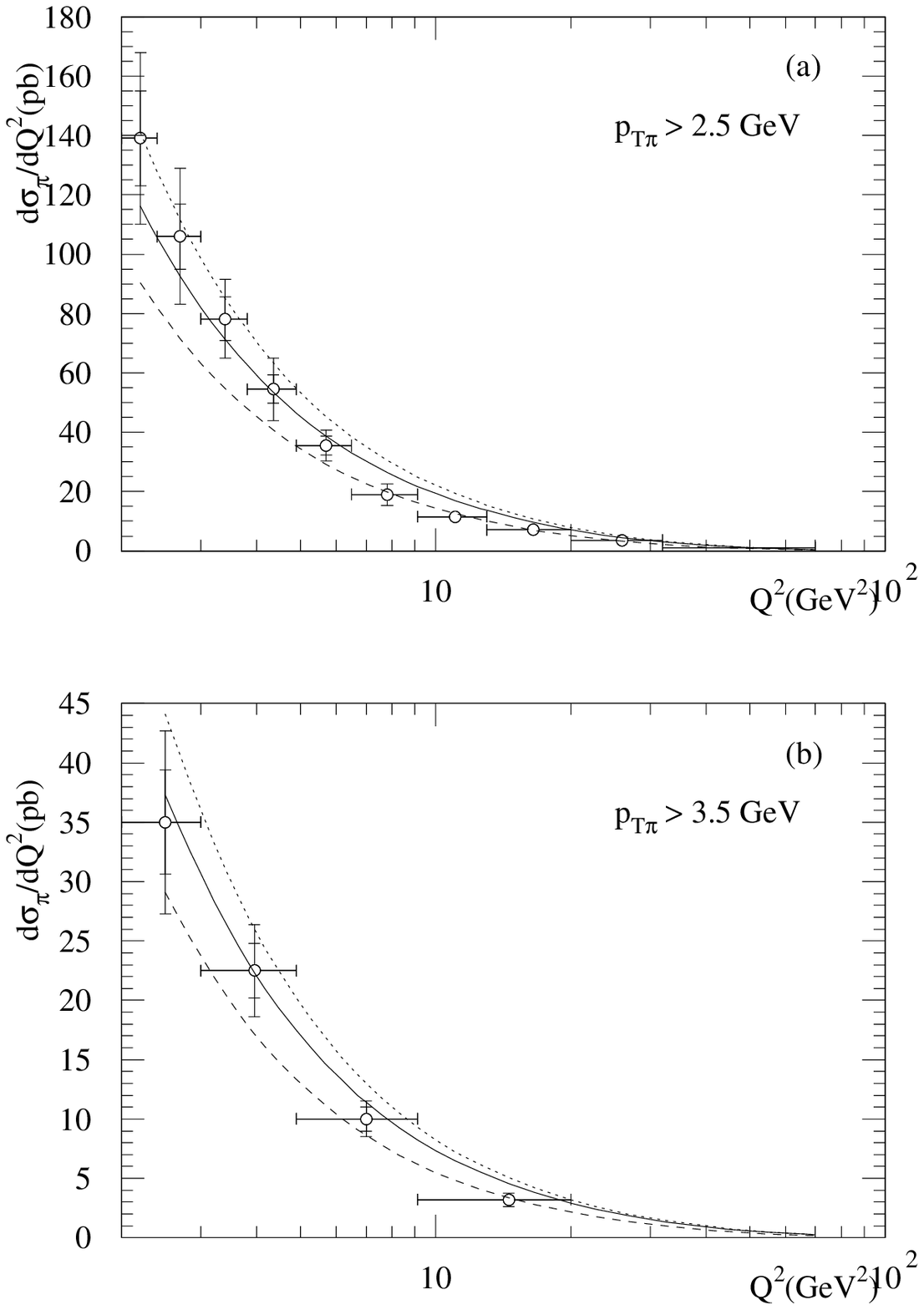,width=12cm}}
\caption{The $\pi^{0}$ differential cross-section versus $Q^{2}$ obtained from 1996 H1 
(preliminary) data \cite{new_h1}.
The curves are predictions, based on the BFKL formalism including sub-leading
corrections, corresponding to the three choices of scales and infrared
cut-off given in (\ref{eqn:scales})
.}
\end{center}
\end{figure}
\begin{figure}
\label{fig:d_etapi0}
\begin{center}
\mbox{\epsfig{figure=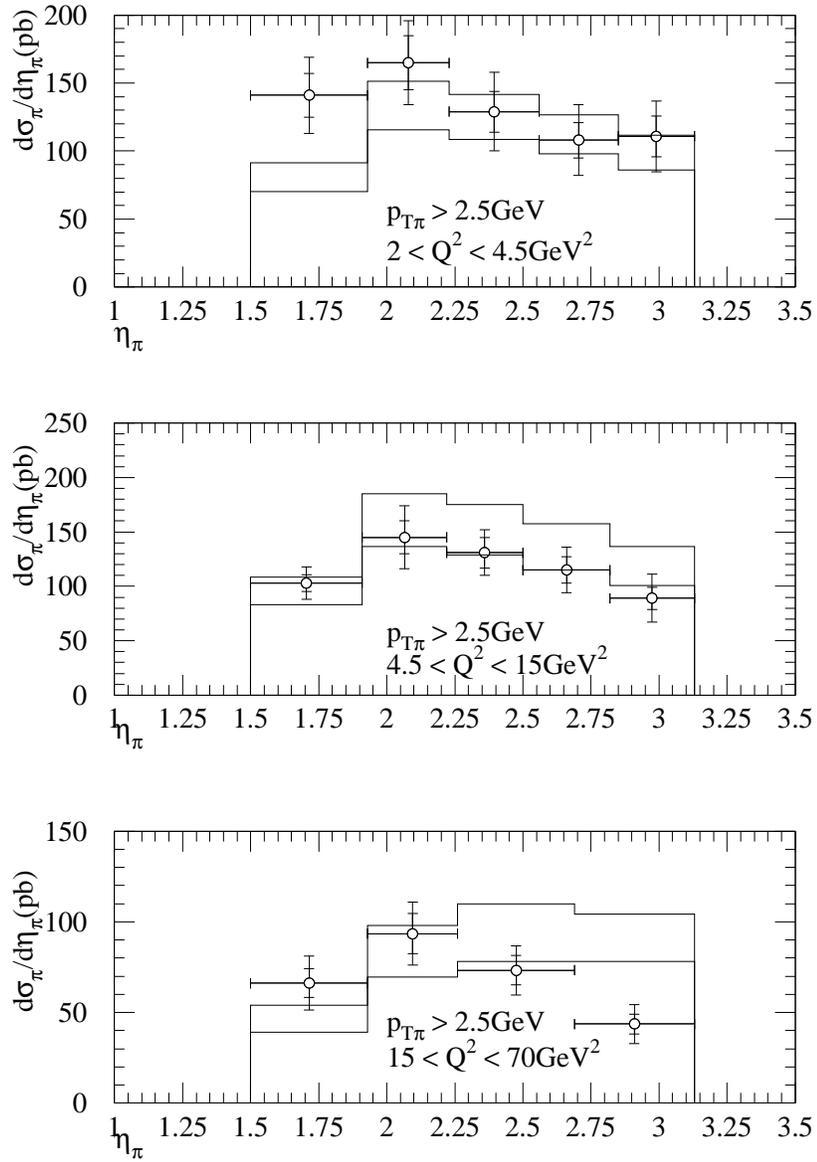,width=12cm}}
\caption{The $\pi^{0}$ differential cross-section versus rapidity obtained from 1996 H1 
(preliminary) data \cite{new_h1}.
The upper and lower histograms are the predictions of the BFKL 
formalism including subleading
corrections respectively corresponding to choices (ii) 
and (iii) of the scales and
$k_{0}^{2}$ given
in (\ref{eqn:scales})
.}
\end{center}
\end{figure}
\section{Forward dijet production}
\begin{figure}
\label{fig:d_2jet}
\begin{center}
\mbox{\epsfig{figure=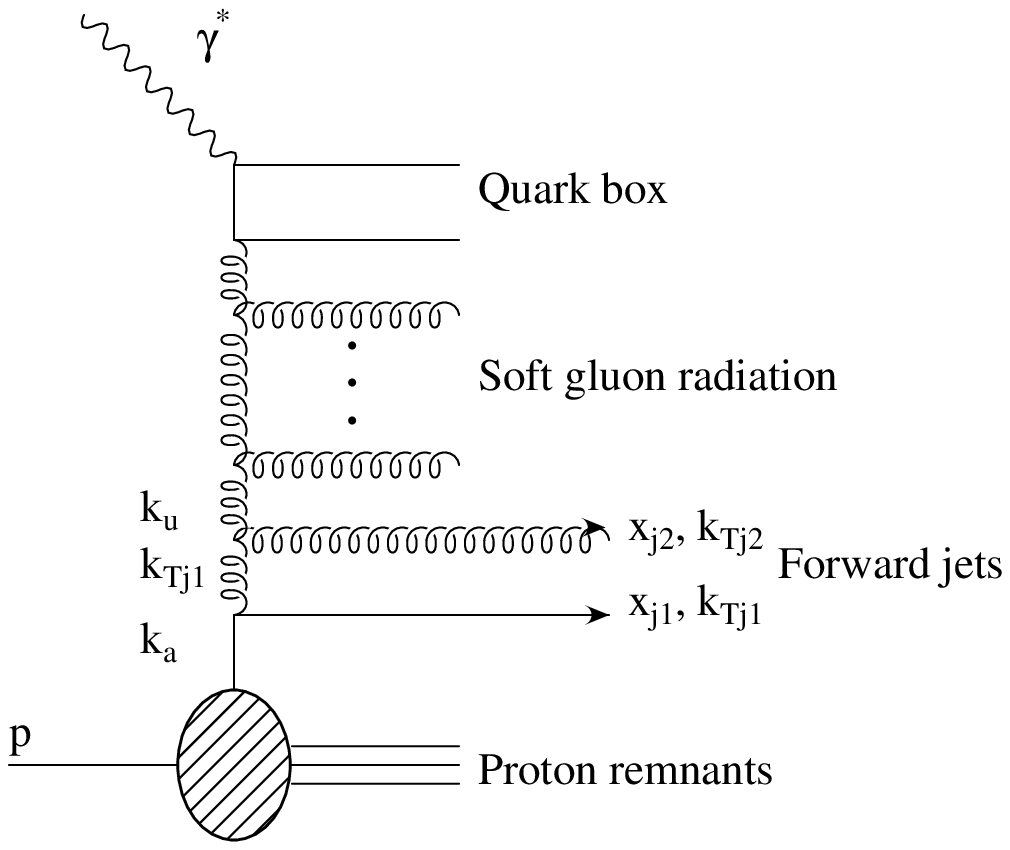,width=10cm}}
\caption{Diagram illustrating the emission of an extra forward jet
in addition to the standard Mueller process.}
\end{center}
\end{figure}
Forward dijet production provides a further complementary 
measure of small $x$ dynamics \cite{claire}
\begin{displaymath}
\gamma^{*} + p \longrightarrow \mathrm{jet}_{1} + \mathrm{jet}_{2} + {\it X}
\end{displaymath}
The additional second jet is then required to fulfil the same forward 
experimental cuts as for the simple jet data.
The process is  shown in Fig. 11.
We may neglect the effects of soft gluon radiation in the rapidity gap, 
as the experimental constraints are too restrictive to allow significant development 
of a second chain of gluon emission between the jets \cite{claire}
\begin{displaymath}
x_{j2} \lower .7ex\hbox{$\;\stackrel{\textstyle<}{\sim}\;$}x_{j1} \sim O (1).
\end{displaymath}
Analogously to the single forward jet case, the dijet cross-section 
is determined by
\begin{displaymath}
{ \partial \sigma \over \partial x \partial Q^{2} \partial x_{j1}
\partial k_{jT1}^{2} \partial x_{j2} \partial k_{jT2}^{2} } =
{4 \pi \alpha^{2} \over x Q^{4} }
\left[ (1-y) { \partial F_{2} \over \partial x_{j1}
\partial k_{jT1}^{2} \partial x_{j2} \partial k_{jT2}^{2} } +
{1\over 2}y^{2} { \partial F_{T} \over \partial x_{j1}
\partial k_{jT1}^{2} \partial x_{j2} \partial k_{jT2}^{2} } \right],
\end{displaymath}
\begin{equation}
\end{equation}
with
\begin{equation}
x_{j2} { \partial F_{i} \over \partial x_{j1}
\partial k_{jT1}^{2} \partial x_{j2} \partial k_{jT2}^{2} } =
{1\over 2\pi} \int_{0}^{2\pi}d\phi
\Phi_{i}(x/x_{j2},k_{u}^{2},Q^{2}) {\bar{\alpha_{S}}
\bar{\alpha_{S}} \over k_{u}^{2} k_{jT2}^{2} k_{jT1}^{2} }
\sum_{a} f_{a}(x_{j1},k_{jT1}^{2}).
\label{eqn:2jet}
\end{equation}
Here, $\phi$ is the azimuthal angular separation between the 
two forward jets, and $k_{u}^{2}$ is given by
\begin{displaymath}
k_{u}^{2} = {(\mbox{\boldmath${k}_{jT1}$} + \mbox{\boldmath${k}_{jT2}$})}^{2}.
\end{displaymath}
The azimuthal ($\phi$) integration is performed numerically.
Experimental jet resolution effects are included by imposing a cut, 
$\mathrm{{\it R}_{min}}$ in 
rapidity-azimuthal angle space:
\begin{displaymath}
\sqrt{ (\Delta\eta)^{2} + (\Delta\phi)^{2} } > \mathrm{{\it R}_{min}} = 1.
\end{displaymath}
The H1 results published in 1994 quote a result of $6.0 \pm 0.8$
(stat) $\pm 3.2$(syst) pb for the total cross-section of DIS + 2 forward jet
events \cite{h1}.
The calculation, including the consistency constraint, and imposing 
all the experimental cuts imposed by the H1 collaboration, results in 
a cross-section of $5.2, 4.8$ or $2.7$pb depending on the different scales in 
$\alpha_{S}$ in (\ref{eqn:2jet}) chosen as in (\ref{eqn:scales}) respectively. 
The predictions for the two-jet/one-jet ratio give $1.0, 1.1$ and $0.8$\%
respectively, to be compared with the H1 measurement of $1.1 \pm 0.6$\%.
That is, small $x$ QCD is able to satisfactorily describe the observed rate
of forward dijet production.
\section{Discussion}
\begin{figure}
\label{fig:vary}
\begin{center}
\mbox{\epsfig{figure=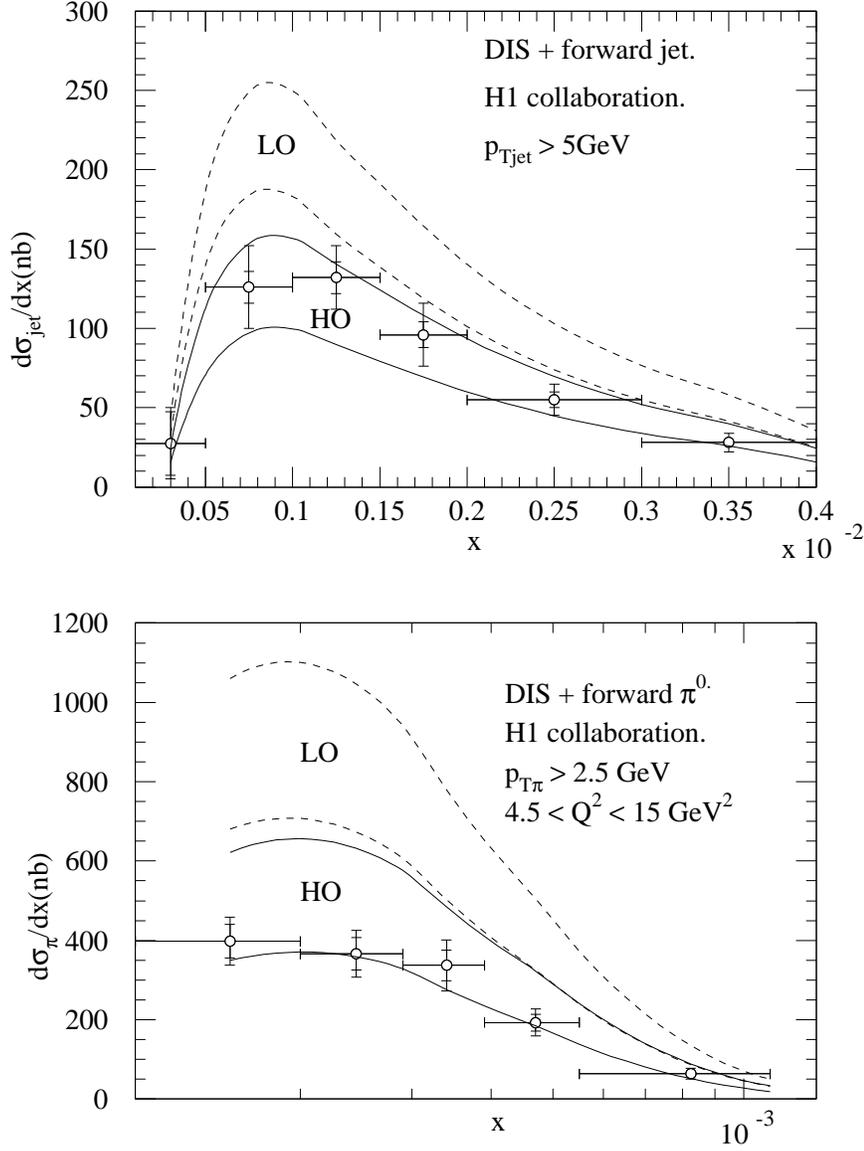,width=12cm}}
\caption{
The differential cross-section for (a) DIS + forward jet, 
(b) DIS + forward $\pi^{0}$ (1996 H1 preliminary), events. 
The continuous and dashed curves
correspond to the inclusion and omission of the consistency constraint.
The lower curve of each pair corresponds to choosing the scales of 
$\alpha_{S}$ in (\ref{eqn:qbox}) and (\ref{eqn:partons}) to be
$k_{T}^{2} + m_{q}^{2}$ and $k_{T}^{2}$ respectively. The upper curves
have scales $1/4$ of these values. In all cases the infrared cut-off in
(\ref{eqn:mbfkl}) is taken to be $k_{0}^{2} = 0.5$GeV$^{2}$.
The jet data are at the hadron level; converting to the parton level
is expected to lower the jet data points by about 15-20\% \cite{wengler_pc}.
}
\end{center}
\end{figure}
The process of deep inelastic scattering (DIS) with a forward jet has quite a
chequered history. The original ``gold-plated'' proposal of using DIS $(x,Q^{2})$
events containing a forward $(x_{j}, k_{jT}^{2})$ jet to study small $x$
dynamics originally dates back to Mueller \cite{mueller}. The idea
was to study small $x/x_{j}$ dynamics with $k_{jT}^{2} \sim Q^{2}$ so
that DGLAP effects are absent. 
The observation of a forward jet allows the deep inelastic scattering to take
place off a parton (in a domain where the distribution
$f_{a}(x_{j}, k_{Tj}^{2})$ is well known) so that the small $x/x_{j}$
dynamics can be studied free from the uncertainties in proton structure.
The detailed formalism was provided
soon after \cite{sutton2}. Then it was the turn of the experiments to collect a
DIS + forward jet data sample. To obtain sufficient statistics, jets of relatively
low $k_{jT}^{2}$ had to be considered, bringing problems of jet identification
and measurement. On the theoretical side all phenmenological analyses
\cite{sutton2,bartels,sabine1} 
were based on the LO BFKL formalism, with a tendency of the 
predictions to overshoot the observed DIS + forward jet cross-section.
At the same time it was also shown that one cannot obtain a sufficient increase
of the cross-section, with decreasing $x$, from fixed order QCD calculations.
In fact the fixed order (NLO) predictions are found to be a factor of 4
or more below the data \cite{bartels,mirkes}.\\
Recently a different approach \cite{kramer,jung} to forward jet 
production has been proposed,
based on describing
the data in terms of the partonic structure of the virtual photon and DGLAP 
evolution.
In this picture the dominant contribution in the region $Q^{2} < k_{Tj}^{2}$
to the production of forward jets comes from the hard scattering of partons 
in the virtual photon with those in a proton. The calculations presented in
ref. \cite{jung} are based on the parton 
distribution functions of the virtual 
photon evolved from some (low) scale $Q_{0}^{2}$ to the scale $\mu^{2} = -\hat{t}$
(which is greater than $k_{Tj}^{2}$ but less than $k_{Tj}^{2} + Q^{2}$). 
The parton distributions in
virtual photons which are used in the estimates of the forward jet 
cross-sections were taken from ref. \cite{sas}. These parton distributions
contain a rather arbitrary, although phenomenologically plausible, 
parameterization of the $Q^{2}$ dependence.
At large $Q^{2}$, however, the parton distributions in the virtual photon are dominated
by the point-like contribution which can be completely specified in
perturbative QCD. In fact the approach developed in \cite{kramer} is not 
very sensitive to the specific parameterisation of the parton distributions of the
virtual photon.
Our calculations based upon the BFKL equation may also be interpreted in
terms of hard scattering of partons in a virtual photon. 
In this case, however, the $Q^{2}$
dependence of the parton distributions is not arbitrary, but rather is dynamically
specified by the impact factor defined by the quark box.
This impact factor corresponds to the point-like $\gamma^{*}q\bar{q}$ coupling.
The analogy with the resolved photon picture becomes more evident if we
approximate the BFKL evolution by its double leading logarithm approximation
corresponding to strongly-ordered transverse momenta from $Q^{2}$ towards the
hard scale $k_{Tj}^{2}$. The BFKL-based calculation should not therefore be
interpreted as an alternative explanation to the resolved virtual photon
picture since the latter is just part of the former. 
The BFKL description also has the merit of treating in a unified way all
possible kinematical configurations. It does not in particular divide
the underlying mechanism depending upon the relations between the 
potentially large scales ($Q^{2}$ and $k_{Tj}^{2}$) of the problem.\\
There have been two recent developments, both of which we have considered in
this paper. First, DIS + forward $\pi^{0}$ process has also been measured. 
These data provide a 
complementary measurement which overcomes the experimental ambiguities inherent
in the measurements of forward jets, albeit at the expense of a reduced rate.
Second, as far as BFKL theory itself is concerned, there now exist complete
NLO results available. These indicate that sub-leading effects are very important
and cannot be neglected in a confrontation of small $x$ dynamics with the
data. In this paper we include subleading contributions in the description
of DIS + forward jet and forward $\pi^{0}$ data. 
The effect of including the subleading terms can be seen from Fig. 12.
We show the predictions for DIS + forward jet and DIS + forward $\pi^{0}$
with and without the consistency constraint included for two physically
reasonable choices of scale. We see that the subleading terms (that is the
consistency constraint) reduce the predictions at the smaller values of $x$ by 
almost a factor of two. There is a sizeable ambiguity in the predictions
due to the choice of scale, but nevertheless both sets of data favour the 
inclusion of the sub-leading terms. 
Even though
the subleading effect suppresses the cross-sections, the predictions
remain sufficiently 
steep with decreasing $x$ to describe the data. 
By inspecting the comparison shown in the two plots in Fig. 12 we note 
that, relative to the data, the DIS + forward $\pi^{0}$ predictions are
lower than those for the DIS + jet process. 
However the comparison for the DIS + $\pi^{0}$ process 
is $Q^{2}$ dependent, see Fig. 9 and 7.
On average the forward $\pi^{0}$ predictions are about 20\% above the
data if the scales are chosen so as to give an optimum description
of the forward jet data. This discrepancy is well within the
total uncertainties.
Moreover, as was mentioned before, 
the experimental data on the forward jet cross-sections
correspond to the measurements at the hadron level, while our theoretical
predictions concern the parton level. 
The hadronisation effects are expected to raise 
the parton level jet cross-section by about 15-20\% \cite{wengler_pc}, and so the overall 
consistency between the forward jet and $\pi^{0}$ data and the small $x$ QCD
predictions shown at the parton level in Fig. 12, is in fact better than shown.
Since we can identify and measure less energetic forward $\pi^{0}$'s
than jets, we are able to sample smaller values of $x$ in the former process.
We conclude that there exists an economical, 
physically-based description of both the DIS + forward jet and the 
DIS + forward $\pi^{0}$ data in terms of the small $x$ QCD framework.
\section*{Acknowledgements}
We thank Franz Eisele, Thorsten Wengler, Ewelina Mroczko-\L{}obodzi\'{n}ska
and Jesus Guillermo Contreras for their interest in this work 
and for information concerning the data. 
We thank Marco Stratmann and Mark W\"{u}sthoff for useful discussions,
and Bjorn P\"{o}tter for illuminating comments. 
JK thanks the Physics Department and Grey College of the University of 
Durham, and JJO thanks the Institute for Nuclear Physics of Krakow
for their warm hospitalities.
This work was supported in part 
by the UK Particle Physics and Astronomy Research Council, by the Polish 
State Committee for Scientific Research (grant no. 2 P03B 089 13) and
by the EU Fourth Framework Programme TMR, Network `QCD and Particle
Structure' (contract FMRX-CT98-0194, DG12-MIHT).

\end{document}